# OpenCitations, an infrastructure organization for open scholarship

**Authors**

Silvio Peroni – https://orcid.org/0000-0003-0530-4305
silvio.peroni@opencitations.net; silvio.peroni@unibo.it
Digital Humanities Advanced Research Centre, Department of Classical Philology and Italian Studies, University of Bologna, Bologna, Italy
Research Centre for Open Scholarly Metadata, Department of Classical Philology and Italian Studies, University of Bologna, Bologna, Italy

David Shotton – https://orcid.org/0000-0001-5506-523X
david.shotton@opencitations.net; david.shotton@oerc.ox.ac.uk
Oxford e-Research Centre, University of Oxford, Oxford, United Kingdom
Research Centre for Open Scholarly Metadata, Department of Classical Philology and Italian Studies, University of Bologna, Bologna, Italy

**Abstract**

OpenCitations is an infrastructure organization for open scholarship dedicated to the publication of open citation data as Linked Open Data using Semantic Web technologies, thereby providing a disruptive alternative to traditional proprietary citation indexes. Open citation data are valuable for bibliometric analysis, increasing the reproducibility of large-scale analyses by enabling publication of the source data.

Following brief introductions to the development and benefits of open scholarship and to Semantic Web technologies, this paper describes OpenCitations and its datasets, tools, services and activities. These include the OpenCitations Data Model; the SPAR (Semantic Publishing and Referencing) Ontologies; OpenCitations' open software of generic applicability for searching, browsing and providing REST APIs over RDF triplestores; Open Citation Identifiers (OCIs) and the OpenCitations OCI Resolution Service; the OpenCitations Corpus (OCC), a database of open downloadable bibliographic and citation data made available in RDF under a Creative Commons public domain dedication; and the OpenCitations Indexes of open citation data, of which the first and largest is COCI, the OpenCitations Index of Crossref Open DOI-to-DOI Citations, which currently contains over 445 million bibliographic citations and is receiving considerable usage by the scholarly community.

# 1. Introduction

Bibliographic citations – the conceptual directional links from a citing bibliographic entity to a cited bibliographic entity created when the author of a published work acknowledges other works in its bibliographic references (Peroni & Shotton 2018a) – are one of the most fundamental types of bibliographic metadata, and are central to the world of scholarship. They knit together independent works of scholarship into a global endeavour, and are important for assigning credit to other researchers. The open availability of citation data is a crucial requirement for the bibliometrics and scientometrics domain, since it "is essential to promote reproducibility and appraisal of research, reduce misconduct, and ensure equitable access to and participation in science" (Sugimoto et al. 2017).

At present, the two most authoritative sources of citation data are Clarivate Analytics' Web of Science (WoS), which grew from the Science Citation Index created by Eugene Garfield in 1964, and Elsevier's Scopus, launched in 2004. Neither are open, most research universities having to pay tens of thousands of dollars annually to access one or both of them, while institutions and independent scholars that cannot afford such costs have no access. More recently, in addition to a number of subject-specific indexes, other sources of general citation data have been made available by other commercial companies, for example by Google (Google Scholar), Digital Science (Dimensions) and Microsoft (Microsoft Academic, formerly Microsoft Academic Search). However, while access to these is free, all have license restrictions on users' ability to reuse and republish the citation data they provide, which limits the full description and reproducibility of research studies using these data.

OpenCitations (http://opencitations.net) has been established for the specific purpose of disrupting that status quo, by providing a fully free and open alternative for accessing global scholarly citation data in Linked Data form (Bizer, C., Heath, T., & Berners-Lee 2009). In this article we introduce the main data and services that OpenCitations provides, we describes the known uses of its data within the scientometrics community, and we describe the planned future developments in terms of new data and initiatives.

An additional excellent source of open citation data, albeit limited to the biomedical domain, is the recently released NIH Open Citation Collection, published by the Office for Portfolio Analysis of the Office of the Director of the U.S. National Institutes of Health (Hutchins et al. 2019). Citations from this resource, which are specifically citations between articles indexed in PubMed, can be retrieved using the iCite web service at https://icite.od.nih.gov.

# 2. From the origins to the Initiative for Open Citations

The first dataset released which contained open bibliographic and citation data was the OpenCitations Corpus, first made available in 2010 as the main output of the Open Citations

Project funded by JISC (Shotton 2013b). JISC (the Joint Information Systems Committee, now Jisc, a United Kingdom not-for-profit company) which was at that time a UK government-funded organization providing infrastructure services such as JANET, a high-speed network for the UK research and education community (https://www.jisc.ac.uk/janet), and awarding grants to support the development of innovative digital solutions for UK education and research. However, this initial version of the OpenCitations Corpus was of limited scope. Between 2010 and 2016, a small number of reference lists had also been openly available by Crossref (https://crossref.org). However the availability of open references from that source changed drastically in April 2017 as a result of the Initiative for Open Citations (I4OC, https://i4oc.org).

The Initiative for Open Citations – of which OpenCitations is one of the founding members, together with the Wikimedia Foundation, PLoS, eLife, DataCite, and the Centre for Culture and Technologies at Curtin University – was created to promote the release of open citation data, specifically by asking scholarly publishers, who were already depositing the reference lists of their publications at Crossref, to make them open to everyone. Before I4OC started, just 1% of all the references deposited at Crossref by scholarly publishers were open. Following discussions between I4OC and the major publishers, and other events publicizing the Initiative over the past two years, more than 1200 scholarly publishers (as of September 2019), including almost all the major scholarly publishers, have chosen to open their deposited references at Crossref. As a result, the percentage of papers with references deposited at Crossref for which the references are open has risen from 1% to 59%, and there are now over 500 million references openly available via the Crossref API. These major publishers include the American Geophysical Union, the Association for Computing Machinery, BMJ, Cambridge University Press, Cold Spring Harbor Laboratory Press, EMBO Press, IOP Publishing, MIT Press, Oxford University Press, the Royal Society of Chemistry, SAGE Publishing, Springer Nature, Taylor & Francis, and Wiley.

The Initiative has also gathered an impressive group of important and supportive stakeholders, including libraries (including the California Digital Library and the British Library), consortia (including the Open Research Funders Group and the Open Access Scholarly Publishers Association), projects (including CORE and OpenAIRE), organizations (including Mozilla and the International Society for Scientometrics and Informetrics), companies (including Microsoft Research and Figshare), and, in particular, funders (including the Bill & Melinda Gates Foundation and the Wellcome Trust) (Shotton 2018). In addition, several international events, including the 2018 Workshop on Open Citations (https://workshop-oc.github.io) and WikiCite 2018 (https://meta.wikimedia.org/wiki/WikiCite_2018) have been held to promote the open availability of citation data.

These open Crossref references have been the basis for the creation and/or extension of commercial services such as ScienceOpen (https://www.scienceopen.com) and Dimensions (https://www.dimensions.ai), and OpenCitations has also used them to populate COCI, the OpenCitations Index of Crossref open DOI-to-DOI citations, described below. In addition, several organizations are continuously acting to increase the support and provision of open citation data. For instance, OASPA (the Open Access Scholarly Publishers Association) has recently introduced the requirement that its members who are depositing reference lists with Crossref must make these reference lists openly available, in line with I4OC recommendations (see https://oaspa.org/oaspa-and-the-initiative-for-open-citations/).

# 3. The benefits of open scholarship and open citations

The Initiative for Open Citations is just one of the disruptive changes that are presently characterising the scholarly digital publishing landscape. As the open scholarship model gains traction, subscription models for access to journal content are crumbling, as evidenced by the Big Deal Cancellation Tracking website maintained by SPARC (https://sparcopen.org/our-work/big-deal-cancellation-tracking/), which presently lists in excess of fifty academic libraries, institutions and consortia that have cancelled their subscriptions to the journals of Elsevier and other major publishers, judging that they are no longer value for money.

The past two decades have seen an explosive growth of open scholarly information sources, providing radical alternatives to commercial sources of scholarly information. Predicated upon the belief that all the fruits of publicly funded scholarly endeavour should be openly available to all scholars and to the general public, and complementing the open publication of knowledge more generally by Wikipedia (https://www.wikipedia.org/) and similar sources, the open scholarship (open science) movement can be characterized as having four distinct phases:

- **Open source software**, of which Linux as an operating system (https://www.linux.com/) and the Free Software Foundation (https://www.fsf.org/) are good examples.
- **Open access publication** of scientific articles and the rise of Open Access publishers, among which the Public Library of Science (PLoS; https://www.plos.org/), eLife (https://elifesciences.org/) and F1000 Research (https://f1000research.com/) are prominent.
- **Open research datasets**, held in repositories of which the Protein Databank (https://www.wwpdb.org/), and the Dryad Data Repository (https://datadryad.org/), are just two prime examples among many in the biological realm, and espoused by organizations such as CODATA (http://www.codata.org/), the Research Data Alliance (https://www.rd-alliance.org/) and the European Open Science Cloud (EOSC, https://www.eosc-portal.eu/).
- **Open bibliographic metadata**, held by scholarly infrastructure organizations such as PubMed (https://www.ncbi.nlm.nih.gov/pubmed) and Crossref (https://www.crossref.org/), whose services are invaluable to scholars worldwide.

Bibliographic metadata, the information describing scholarly publications, are key components of the open scholarly ecosystem, and prime among these are bibliographic citations, described above in the Introduction. Bibliographic citations are factual in nature, and such facts cannot be copyrighted and should not be placed behind subscription paywalls.

Those that benefit from open citations include:

- **Researchers and authors**, particularly those who are not members of the elite club of research universities that can afford subscription access to the commercial citation indexes Web of Science and Scopus.
- **Bibliometricians**, who will be able to publish the research data upon which their research findings are based.
- **Librarians**, who will be better able to support their stakeholder communities (authors, researchers, students, institutional administrators) by providing free access to citations.
- **Funders**, who can better assess the impact of scientific work and decide which researchers, ideas and projects are worth funding.
- **Academic administrators of research institutes and universities**, who will more easily be able to track the scholarly productivity and influence of their members.
- **Research managers**, who will have open citation data available for integration in their CRIS systems, including those using CERIF, the Common European Research Information Framework.
- **Data repositories**, that will benefit from open bibliographic citations between their datasets and the articles describing them.
- **Publishers**, who will have more readers guided from open citation data to their online journals, which will secondarily attract additional article submissions.
- Finally, **computer scientists and software providers**, who will be able to exploit this free availability of citation data to build new applications and visualizations that we cannot even begin to imagine.

# 4. A brief introduction to Semantic Web technologies

The change towards open scholarship described in the previous section is not only social but also technological. Recently, one of the most discussed issues has been how to create a scalable infrastructure that can store and serve all the bibliographic and citation data of interest. A centralised solution, i.e. having all the information stored in a unique database, is not feasible in the long term, as the amount of data to be handled increases, due to performance issues and physical space needed for managing a service (access, REST API, query) appropriately. The Confederation of Open Access Repositories (COAR) has recently discussed this specific aspect in its 2017 report (COAR WG Next Generation Repositories 2017). It suggests that there is an urgent need of a "distributed globally networked infrastructure for scholarly communication", which should enable "cross-repository connections" by using "bi-directional links", and that should provide its data using formats that are "machine-friendly, enabling the development of a wider range of global repository services". Semantic Web technologies are one of the strongest options for implementing such a globally distributed scholarly infrastructure.

The World Wide Web Consortium (W3C, https://w3.org) has designed a suite of Semantic Web technologies and standards (https://www.w3.org/standards/semanticweb/) that facilitate the use of precise semantics for the encoding and making machine-processable data on the Web.

The main characteristics of the Semantic Web are very simple:

- All entities (e.g. this article and its authors), their types (e.g. the classes to which all documents and people belong, for example "journal article", "author"), and their relationships (properties defining attributes or linking entities and their types together according to a particular semantics, e.g. the fact that this article *is authored by* Silvio Peroni and David Shotton), are identified by HTTP URIs (https://en.wikipedia.org/wiki/Uniform_Resource_Identifier), and their related information can be retrieved from the Web by using such URIs.
- Publicly available and commonly accepted structured vocabularies (ontologies, https://en.wikipedia.org/wiki/Ontology_(information_science)) are made available on the Web to define the meanings of the different classes of entities and the relationships that link them.
- The abstract data model used to define the information related to such entities, types, and relationships is the Resource Description Framework (RDF) (Cyganiak, Wood, & Krötzsch, 2014).
- Each statement related to a particular entity is expressed in RDF as a subject – predicate – object 'triple', for example `<Paper A> cito:cites <Paper B>`, where CiTO (http://purl.org/spar/cito) is the ontology in which the property "cites" is defined.
- RDF statements stored in different sources can be combined into interconnected information networks (directed graphs, https://en.wikipedia.org/wiki/Directed_graph) – forming 'linked data' – thereby creating a web of knowledge, the Semantic Web, in which the truth content of each original statement is maintained.
- RDF triples can be stored in a particular type of database designed for graph data, known as a triplestore (https://en.wikipedia.org/wiki/Triplestore).
- Such data can be queried using SPARQL (Harris & Seaborne, 2013), a special query language able to retrieve and manipulate data stored in RDF format.

Further explanations of the Semantic Web and its applications to bibliographic information are to be found in a series of short didactic blog posts entitled "Libraries and linked data" to be found on our Semantic Publishing blog (Shotton 2013a).

# 5. OpenCitations: tools and services

OpenCitations is an independent infrastructure organization for open scholarship dedicated to the publication of open bibliographic and citation data by the use of Semantic Web (Linked Data) technologies. It is also engaged in advocacy for open citations, particularly in its role as a key founding member of I4OC. For administrative convenience, OpenCitations is managed by the separate newly formed Research Centre for Open Scholarly Metadata at the University of Bologna (https://openscholarlymetadata.org), of which the two authors of this article, Silvio Peroni and David Shotton, are Director and Associate Director, respectively.

OpenCitations espouses fully the founding principles of Open Science. It complies with the FAIR data principles (Wilkinson et al., 2016) proposed by Force11 (https://www.force11.org) that data should be **findable**, **accessible**, **interoperable** and **re-usable**, and it complies with the recommendations of I4OC that citation data in particular should be **structured**, **separable**, and **open**. On the latter topic, OpenCitations has recently published a formal definition of an Open Citation (Peroni & Shotton, 2018a), and has launched a system for globally unique and persistent identifiers (PIDs) for bibliographic citations – Open Citation Identifiers (OCIs).

## 5.1. Open Citation Identifiers

The Open Citation Identifier (OCI) is a globally unique persistent identifier (PID) for open bibliographic citations, as defined by Peroni & Shotton (2018a). The Open Citation Identifier system has been developed by OpenCitations, which maintains a resolution service for OCIs at http://opencitations.net/oci. Given a valid OCI as input, this resolution service is able to retrieve metadata about the identified citation in RDF (either as RDF/XML, Turtle or JSON-LD), or in Scholix, JSON or CSV formats.

Each OCI has a simple structure: the lower-case letters "oci" followed by a colon, followed by two sequences of numerals separated by a dash. The first sequence identifies the citing publication and the supplier database within which its metadata are to be found, e.g. Crossref or the OpenCitations Corpus (OCC), while the second sequence identifies the cited publication and the supplier database within which its metadata are to be found. Each supplier database is defined by a numerical prefix assigned by OpenCitations comprising a short sequence of positive integers delimited by zeros: for example 020 is the Crossref prefix, while 030 is the OCC prefix.

Thus oci:0302544384-0307295288 is a valid OCI for a citation between two publications recorded in the OCC with internal OCC identifiers 2544384 and 7295288 respectively, while oci:02001010806360107050663080702026306630509-02001010806360107050663080702026 305630301 is a valid OCI for a citation between two publications recorded in Crossref with DOIs doi:10.1186/1756-8722-6-59 and doi:10.1186/1756-8722-5-31 respectively. Details of how alphanumeric identifiers such as DOIs are transformed into the purely numerical sequences used in the OCIs are given by Peroni & Shotton (2019). Since OCIs are designed for machine processing rather than human readability, the length of the numerical strings created by such alphanumeric transformations is not a problem.

The main advantages of treating citations as first-class data entities in their own right, with their own globally unique and persistent citation identifiers, are that:

1. all the information regarding each citation can be stored in one place, with associated metadata;
2. citations become easier to describe, distinguish, count and process; and

3. if available in aggregate, citations described in this manner are easier to analyze using bibliometric methods, for example to visualize citation networks or to determine how citation time spans vary by discipline.

It should be noted that OCIs are not opaque identifiers, since they explicitly encode directional relationships between identified citing and cited entities, the provenance of the citation, i.e. the database that contains it, and the type of identifiers used in that database to identify the citing and cited entities. The OCIs defining citations between a group of publications that cite one another thus contain all the information required to construct the citation network of these publications.

Open Citation Identifiers (OCIs) have been accepted by the community, being recognized as persistent identifiers for citations by the EU FREYA Project ([https://www.project-freya.eu](https://www.project-freya.eu)) [(Ferguson et al., 2018)](#) and being registered by Identifiers.org ([https://identifiers.org/oci](https://identifiers.org/oci)), which is a PID registry and meta-resolver.

## 5.2. The OpenCitations Data Model and the SPAR Ontologies

To enable the description of bibliographic and citation information in machine-readable terms, OpenCitations provided, maintains and updates the OpenCitations Data Model (OCDM) [(Daquino, Peroni & Shotton, 2019)](#), briefly summarised in [Figure 1](#).

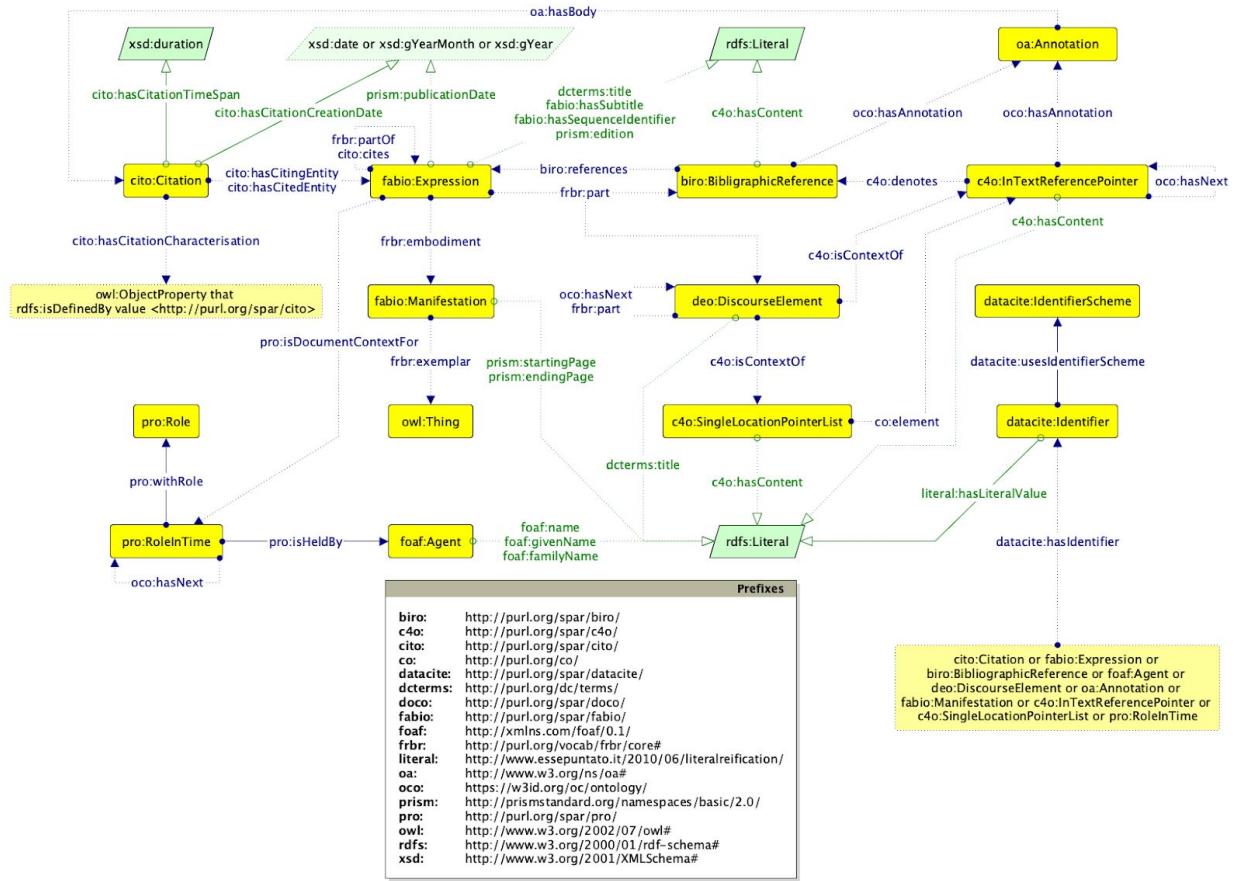

**Figure 1.** A Graffoo diagram summarising the OpenCitations Data Model, implemented using the OpenCitations Ontology (OCO) available at https://w3id.org/oc/ontology.

The OCDM is used to model all the bibliographic and citation entities (i.e. the yellow rectangles in Figure 1, defining the classes of objects the data model allows one to describe), their attributes (i.e. the green arrows) and the relations to other entities (i.e. the blue arrows). All these aspects are exposed in any OpenCitation dataset in RDF, using the 'language' of the Semantic Web, in particular by employing OpenCitations' SPAR (Semantic Publishing and Referencing) Ontologies (http://www.sparontologies.net) (Peroni & Shotton, 2018c). Such usage permits the publication of bibliographic and citation data as Linked Open Data (LOD), thereby conferring machine readability and interoperability of the data on the Web. The OpenCitations Data Model may also be employed by third parties, either for their own use or to structure their data for submission to and publication by OpenCitations.

The OCDM allows one to record information about:

- published *bibliographic resources* (class `fabio:Expression` in Figure 1) that either cite or are cited by another published bibliographic resources, or that contain citing/cited entities (e.g. a journal containing an article or a book containing chapter);

- possible *resource embodiments* (class `fabio:Manifestation` in [Figure 1](#)) defining the particular physical or digital format in which a bibliographic resource was made available;
- *bibliographic references* (class `biro:BibliographicReference` in [Figure 1](#)) usually occurring in the reference list (and usually denoted by one or more in-text reference pointers within a citing bibliographic resource) of a citing entity, that references another bibliographic resource;
- *responsible agents* (class `foaf:Agent` in [Figure 1](#)), such as people or organizations, having a certain role with respect to a bibliographic resource (e.g. an author of a paper or book, or the publisher of a journal);
- the *roles* (class `pro:RoleInTime` in [Figure 1](#)) held by an agent with respect to bibliographic resources (e.g. a person being the author of an article and the editor of another book);
- the *citations* (class `cito:Citation` in [Figure 1](#)) between two bibliographic resources;
- the external *identifiers* (class `datacite:Identifier` in [Figure 1](#)), such as DOI, ORCID, PubMedID, Open Citation Identifier, associated with the bibliographic entities.

In November 2019, a new release of the OCDM was published, revised and extended with additional kinds of entities that enable the description of *in-text reference pointers* (class `c4o:InTextRefefencePointer` in [Figure 1](#)) denoting bibliographic references – i.e. the textual devices (e.g. "[1]" or "Peroni & Shotton 2019") that are embedded in the text of a document within the context of a particular sentence, paragraph or section (which are kinds of *discourse elements*, defined by the class `deo:DiscourseElement` in [Figure 1](#)) – and the citations they instantiate (linked via *annotations*, defined by the class `oa:Annotation` in [Figure 1](#)), accompanied by a description of their *functions*, i.e. the reason why a bibliographic resource is cited [(Teufel, Siddharthan & Tidhar 2006)](#).

## 5.3. The OpenCitations datasets

In terms of data, OpenCitations first developed the OpenCitations Corpus (OCC, [http://opencitations.net/corpus](http://opencitations.net/corpus)) [(Peroni, Shotton & Vitali, 2017)](#), a database of open downloadable bibliographic and citation data recorded in RDF and released under a Creative Commons CC0 public domain waiver, which currently contains information about ~14 million citation links to over 7.5 million cited resources. The current content of the OCC has been mainly derived from biomedical articles within the Open Access Subset of PubMed Central ([https://www.ncbi.nlm.nih.gov/pmc/tools/openftlist/](https://www.ncbi.nlm.nih.gov/pmc/tools/openftlist/)), harvested using the Europe PubMed Central REST API ([https://europepmc.org/RestfulWebService](https://europepmc.org/RestfulWebService)), and contains information described following the OpenCitations Data Model.

In addition and separately, OpenCitations is currently developing a number of Open Citation Indexes ([http://opencitations.net/index](http://opencitations.net/index)), using the data openly available in third-party bibliographic databases. The first and largest of these is COCI, the OpenCitations Index of Crossref open DOI-to-DOI citations ([http://opencitations.net/index/coci](http://opencitations.net/index/coci)) [(Heibi, Peroni & Shotton,](#)

[2019a)](), which presently contains information encoded in RDF on more than 445 million citations, released under a CC0 waiver.

The most recent index is CROCI, the Crowdsourced Open Citation Index ([http://opencitations.net/index/croci](http://opencitations.net/index/croci)) [(Heibi, Peroni & Shotton, 2019b)](), which is designed to host citation data submitted by third parties – authors, editors, scholars – allowing them to upload to OpenCitations citation data that are not elsewhere openly available, for example from their own publications, journals and reference collections, in an effort to fill the gap of missing citations from some publishers (particularly Elsevier) which are not presently available at Crossref as open material.

The Indexes contain information about the citations themselves, in which the citations, instead of being considered as simple links, are treated as first-class data entities in their own right. This permits us to endow each citation with descriptive properties, such as the date on which the citation was created, its timespan (i.e.the interval between the publication date of the cited entity and the publication date of the citing entity), and its type (e.g. whether or not it is a self-citation). An in-depth description of the definition and use of citations as first-class data entities is provided by [Shotton (2018)](). In contrast to the OCC, these Indexes do not store metadata about the citing and cited bibliographic entities internally.  Rather, these entities are identified in the Indexes by their unique identifiers (e.g. DOIs), enabling bibliographic information to be retrieved on-the-fly upon request by means of the related API (see the operation "metadata" at [https://w3id.org/oc/index/api/v1](https://w3id.org/oc/index/api/v1) for additional information).

## 5.4. Provenance information

The provenance of a certain resource concerns "the people, institutions, entities, and activities involved in producing, influencing, or delivering a piece of data or a thing" and it "is crucial in deciding whether information is to be trusted, how it should be integrated with other diverse information sources, and how to give credit to its originators when reusing it" [(Moreau & Missier 2013)]().

All the entities (bibliographic resources, citations, agents involved in the publication process such as authors and publishers, etc.) included in the datasets released by OpenCitations are accompanied by provenance information, so as to keep track of the curatorial activities related to each entity, the curatorial agents involved, and the sources used to obtain such data. In addition, OpenCitations also tracks how the data related to its entities may have changed in time, to allow one to reconstruct the particular description status (or snapshot) of an entity at a specified time. This has been technically implemented by extending the Provenance Ontology [(Lebo, Sahoo & McGuinness 2013)]() with a SPARQL-based construct that has been inspired by existing works on change tracking mechanisms in documents created through word-processors such as Microsoft Word and OpenOffice Writer [(Peroni, Shotton & Vitali 2016)]().

## 5.5. OpenCitations software

All the aforementioned data are made available in RDF (Cyganiak, Wood, & Krötzsch, 2014), this being the main format for expressing structured machine-readable data for the Web, and are stored in specialized graph databases for RDF known as triplestores. OpenCitations provides open source software of generic applicability for searching, browsing, and querying its bibliographic and citation data (https://github.com/opencitations). Thus, programmatic access to the OCC and COCI triplestores may be obtained using queries in SPARQL, the RDF query language (Harris & Seaborne, 2013), via their SPARQL endpoints, in JSON and CSV formats by using the OpenCitations REST API created using RAMOSE (https://github.com/opencitations/ramose), OpenCitations' application for creating REST APIs over SPARQL endpoints (Heibi, Peroni & Shotton, 2019c), or via HTTP requests in different formats (HTML, RDF/XML, Turtle or JSON-LD, via content negotiation). The OpenCitations datasets can also be explored by humans using OSCAR, the OpenCitations RDF Search Application (https://github.com/opencitations/oscar), employing author name, work title (or part thereof) or identifier (DOI, ORCID) as input, and the returned results can then be browsed using OSCAR's associated browse interface, LUCINDA, the OpenCitations RDF Resource Browser (https://github.com/opencitations/lucinda). Metadata for individual bibliographic entities within OCC and citations within COCI can also be accessed via a simple Web form using their individual URIs (e.g. https://w3id.org/oc/corpus/br/1), and downloads of the entire OCC and COCI datasets are possible from dumps made periodically and stored on Figshare (http://opencitations.net/download), so as to support large-scale scientometric analyses using the whole content of the datasets.

# 6. The sustainability of OpenCitations

In order to build an infrastructure that is sustainable in the long term and that follows pure Open Science principles, OpenCitations has adopted in full the Principles for Open Scholarly Infrastructures (Bilder, Lin & Neylon, 2015), which recommend three sets of principles to which open scholarly infrastructures should adhere, under the headings *Insurance*, *Governance* and *Sustainability*.

## 6.1. Insurance

OpenCitations completely fulfills the requirements of Bilder, Lin & Neylon (2015) designed to insure that the work of OpenCitations would survive should OpenCitations itself cease to exist. All the software released by OpenCitations is open, available on GitHub at https://github.com/opencitations and released with the ISC License (https://choosealicense.com/licenses/isc/), which is a very permissive free software license that allows maximum reuse of the software in different contexts, either commercial or non-commercial. All the models (i.e. the OpenCitations Data Model and the SPAR Ontologies) used to describe the bibliographic and citation data provided OpenCitations are made available using a CC-BY license (https://creativecommons.org/licenses/by/4.0/), while all the OpenCitations data are released using the CC0 public domain waiver (https://creativecommons.org/publicdomain/zero/1.0/) so as to maximise their reuse, and are obtainable by a variety of means, as described above.

Indeed, since all OpenCitations software and data are open and recorded using open standards, it is possible even now for third parties to take and re-use the data, or migrate it to new platforms, at any time. Finally, it is worth mentioning that OpenCitations does not hold nor will it seek to obtain any patent for any of its products.

## 6.2. Governance

OpenCitations is engaged with citation data covering the whole spectrum of the scholarly research domain. In addition, all the OpenCitations applications – such as OSCAR, LUCINDA, and RAMOSE (https://github.com/opencitations/ramose), used to developed the aforementioned REST APIs – have been designed to be of generic usefulness, and are made available in a manner that permits their reuse by members of the community in a plethora of different scenarios that need not be related in any way to bibliographic and citation data.

OpenCitations, as an independent infrastructure organization for open scholarship, is currently directed by its two Directors, the authors of this article. For administrative convenience, it is managed by the Research Centre for Open Scholarly Metadata, an independent research centre within the University of Bologna, which is itself, as a public Italian university, a non-profit institution. The Research Centre has an International Board drawn from leaders within the main bibliographic stakeholder communities of relevance (librarians, bibliometricians, academics, data service providers, etc.) who have shown past solid commitment to open scholarship. The statutes of the Research Centre will ensure that OpenCitations' original aim of free provision of open bibliographic and citation data, services and software is maintained, and that OpenCitations as an organization cannot in future be taken over or controlled by commercial interests, nor become involved in political, regulatory, legislative or financial lobbying of any kind.

## 6.3. Sustainability

So far, OpenCitations and its products has been funded by specific grants from different investors. In particular, the first OpenCitations Corpus prototype was funded by a small one-year grant from JISC in 2010, which was awarded a six-month extension. There then followed a period without external funding, that ended with the award of a grant from the Alfred P. Sloan Foundation for The OpenCitations Enhancement Project (https://sloan.org/grant-detail/8017), funding we have used to achieve the current status of OpenCitations. We now have a new grant from the Wellcome Trust for a one-year project entitled Open Biomedical Citations in Context Corpus,

(https://wellcome.ac.uk/funding/people-and-projects/grants-awarded/open-biomedical-citations-context-corpus), that started on the 1st of July 2019.

OpenCitations will continue to apply for targetted grants for specific projects, either alone or with partners, by participating in H2020 calls and by approaching additional funders and organizations including the Alfred P. Sloan Foundation (https://sloan.org), the Chan-Zuckerberg Initiative (https://chanzuckerberg.com), and the Arcadia Fund (https://www.arcadiafund.org.uk).

Because all OpenCitations' data and services are open, it has nothing to sell or against which to charge membership fees. Furthermore, OpenCitations' basic philosophy that *all* its data and services should be free mitigates against any form of 'freemium' income generation. Long term sustainability for OpenCitations thus requires ongoing support from the scholarly community, for example following the SCOSS (http://scoss.org) and the Invest in Open Infrastructure (https://investinopen.org) models of crowd-sourced support from stakeholders within the open scholarship community, or alternatively by 'adoption' by one or more major scholarly libraries, philanthropists or academic funding agencies, those institutions or individuals thus gaining credit for supporting OpenCitations in conformity with their missions and goals. As such, OpenCitations is no different from any other open infrastructure organization that provides free data services of value, such as PubMed, Europe PubMed Central or the Protein Data Bank.

We are therefore delighted to announce that OpenCitations has been selected by the SCOSS Board for their second round of crowd-funding support, as OpenCitations aligns well with open science goals and is an innovative service; open citation data are important to the community since they have the potential to support change in research assessment, and if successful could be a game changer by challenging established proprietary citation services. Starting in January 2020, SCOSS will invite research organizations, scholarly institutions and funders of all sizes throughout the world to contribute financially in proportion to their size and ability to sustain OpenCitations' operations over the next three years as it transitions into a global scholarly infrastructure organization with a secure financial footing. As Directors of OpenCitations, we invite and encourage you to support OpenCitations through this SCOSS initiative.

If OpenCitations continues to expand its coverage of the scholarly domain as it has been doing (from ~14 million to over 445 million citations in the past year), so as to offer a genuine alternative in terms of coverage to the extensive citation data offered by WoS and Scopus (WoS has over a billion citations), then it stands every chance of attracting financial support from university libraries and other scholarly institutions at a fraction of the cost of their current subscriptions to those commercial citation indexes.

# 7. Usage statistics

In the past year the OpenCitations website, with all its services and pages, has been accessed more than 3.1 million times by more than 68,000 unique visitors (identified by their IP addresses) – we have excluded from these counts all accesses made by automated agents and bots. Specifically, the number of accesses made between April 2018 and March 2019 (inclusive) is shown in Figure 2, that list five main categories of information access services available in the website, these being the direct HTTP access to a particular bibliographic resource and/or citation ("HTTP_CONT_NEG"), the search/browse interfaces ("INTERFACE"), the REST APIs ("API"), SPARQL queries to the endpoints ("SPARQL"), and 'other' (visits to the OpenCitations homepage and other web pages). It is worth mentioning that the APIs were formally introduced in June 2018, following a few internal experiments run in May 2018. They have rapidly become the main service used for querying the citation data available in OpenCitations.

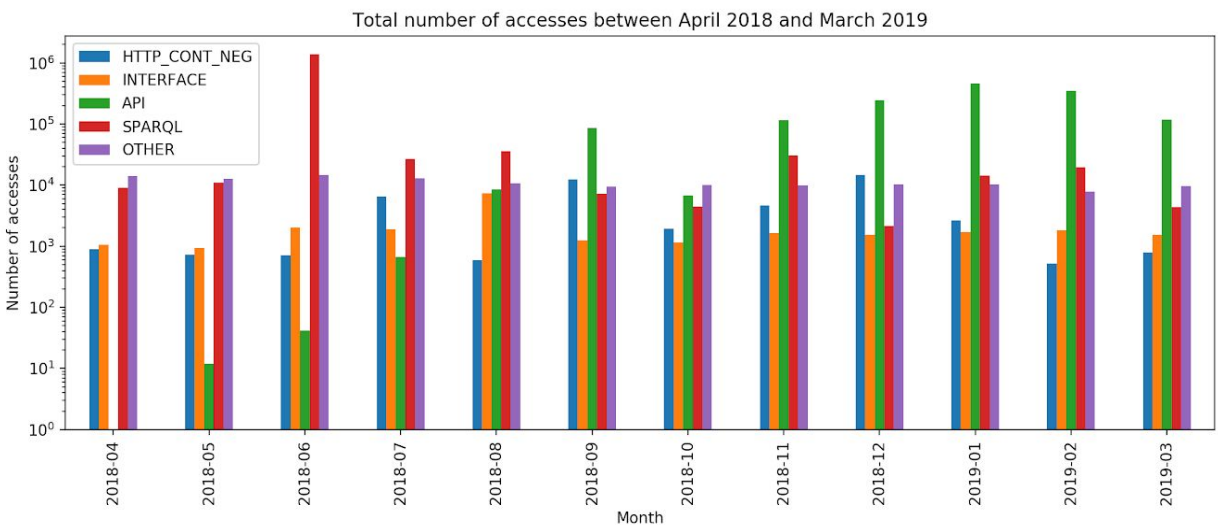

**Figure 2.** The overall number of accesses to the OpenCitations website pages and services in the past year, month by month. "HTTP_CONT_NEG" (i.e. HTTP content negotiation) indicates direct access to stored resources by means of their HTTP URI, "INTERFACE" indicates the use of Web interfaces for browsing and searching bibliographic and citation data, "API" shows the calls to the various OpenCitations REST APIs, "SPARQL" indicates the calls to the OpenCitations SPARQL endpoints, while "OTHER" lists the accesses made to all the other resources. Note that the y-axis is logarithmic.

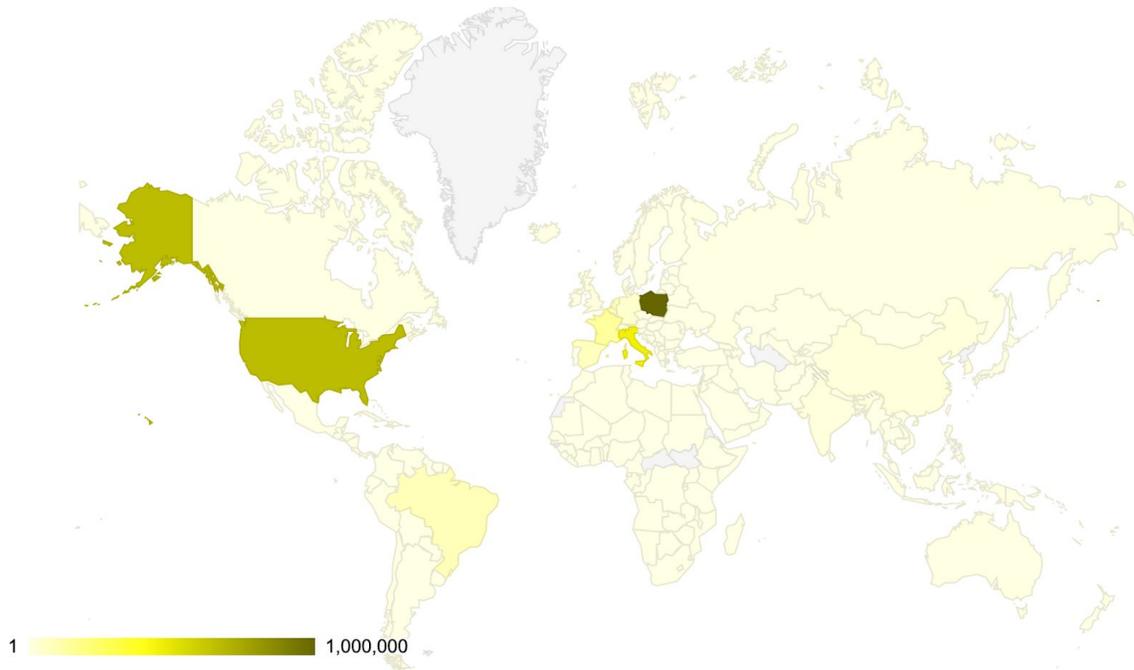

**Figure 3.** The map showing the relative frequency of accesses to the OpenCitations website in the last year, from April 2018 to March 2019, organised per countries. Only a few countries worldwide, coloured in white, did not access the website in the past year.

These data are complemented by the chart shown in Figure 3, which shows the number of requests from different countries worldwide (identified by the IP addresses of the requests). As it is clear from the diagram, those countries making the most requests were Italy, Poland and the United States of America followed by Brazil, France, the Netherlands and Spain, then China, Germany, India and the UK.

In Figure 4, we show the statistics concerning the OpenCitations resources made available on Figshare (i.e. the dumps of the datasets made available by OpenCitations, as well as the definition documents). In the past year, these Figshare documents have obtained more than 20,000 views and 3,000 downloads overall.

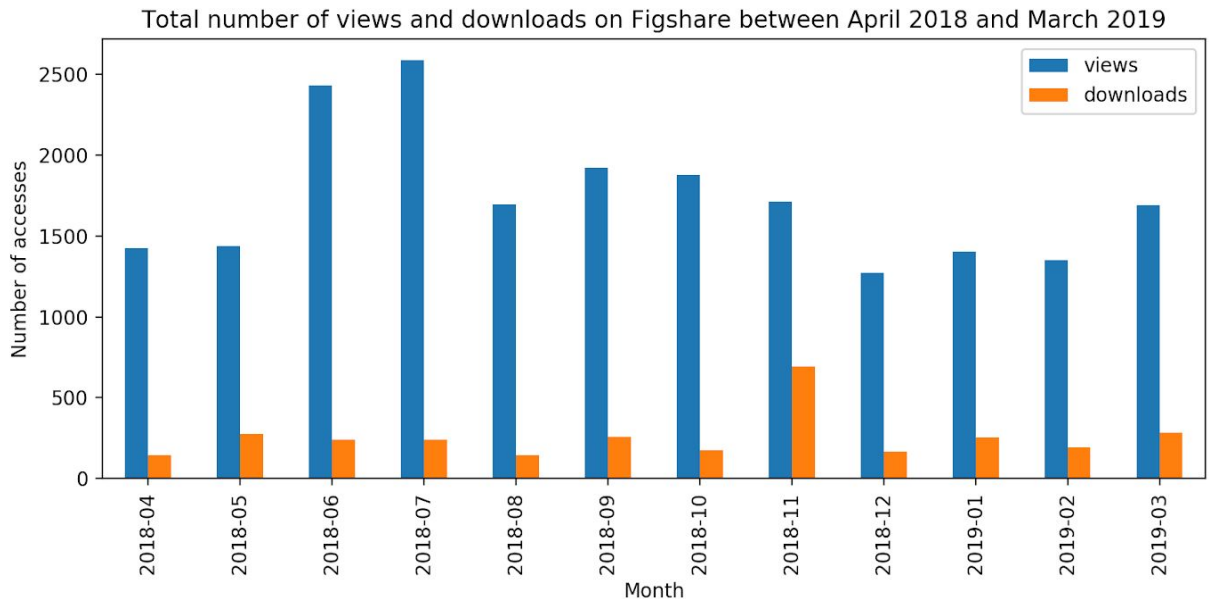

**Figure 4.** The overall number of views and downloads to all the OpenCitations resources stored in Figshare – mainly dataset dumps and definition documents.

All the CSV data used to create the foregoing charts are available in OpenCitations (2019).

## 8. Adoption of OpenCitations by the community

The open citation data published by OpenCitations will benefit all scholars and researchers, particularly those who are not members of the elite club of research universities that can afford subscription access to WoS and Scopus. These data are of particular value to bibliometricians, since they not only permit open research, but also allow re-publication of the actual data upon which the research findings are based, thus enabling the reproducibility of bibliometrics and scientometrics studies. This is rarely possible when the research is based on data from proprietary citation indexes. Such scholars will now be able to pursue their studies with greater freedom, following reference trails through the citation network without hindrance, and have their own publications more easily found, discussed and cited.

In the past months, several researchers have already used the data provided by OpenCitations for such studies. For instance, Kamińska (2017) published a case study showing the possibilities of running bibliometric analysis on the open citation data of PLOS ONE articles available in the OpenCitations Corpus. COCI data, downloaded from the CSV dump available at http://opencitations.net/download, have also been used in at least three bibliometric studies recently. During the LIS Bibliometrics 2019 Event, Pearson (2019) presented a study run on publications by scholars at the University of Manchester which used COCI to retrieve citations between these publications so as to investigate possible cross-discipline and cross-department potential collaborations. Similarly, COCI data were used to conduct an experiment on the latest Italian Scientific Habilitation (the national exercise that evaluates whether a scholar is appropriate to receive an Associate/Full Professorship position in an Italian university), which aimed at trying to replicate part of the outcomes of this evaluation exercise for the Computer Science research field using only *open* scholarly data, including the citations available in COCI, rather than citation data from subscription services (Di Iorio, Peroni, & Poggi, 2019). Finally, COCI has also been used to explore the roles of books in scholarly communication (Zhu et al., 2019).

OpenCitations data have also been used by several tools dedicated to the visualisation of citation graphs and other scholarly networks. VOSviewer (van Eck & Waltman, 2010) (http://www.vosviewer.com) is a software tool, developed at the Leiden University's Centre for Science and Technology Studies (CWTS), for constructing and visualizing bibliometric networks, which may include journals, researchers or individual publications, and may be constructed based on citation, bibliographic coupling, co-citation, and co-authorship relations. Starting from version 1.6.10 (released on January 10, 2019), VOSviewer can now directly use citation data stored in COCI, retrieved by means of the COCI REST API. Citation Gecko (http://citationgecko.com) and OCI Graphe (https://dossier-ng.univ-st-etienne.fr/scd/www/oci/OCI_graphe_accueil.html) are two other examples of web tools that allow one to map a research citation network using some initial seed articles. Both of them use COCI data (accessed via the REST API) to generate the citation network shown in the browser. An alternative visualisation tool is VisualBib (http://visualbib.uniud.it/) (Corbatto & Dattolo, 2018), that uses the data in the OCC, among others, to support researchers who wish to create, refine and visualize bibliography from a small set of significant papers or from a restricted number of authors.

In addition to the aforementioned activities, OpenCitations is currently collaborating with a number of academic projects related to the management of bibliographic and citation data, both to promote the use of the OpenCitations Data Model, and to provide a publication venue for the citation data that these projects are liberating from the scholarly literature. Among these, it is particularly worth mentioning the Venice Scholar Index (https://venicescholar.dhlab.epfl.ch), the Linked Open Citations Database (LOC-DB, https://locdb.bib.uni-mannheim.de) (Lauscher et al., 2018), and the EXCITE Project (http://excite.west.uni-koblenz.de) (Hosseini et al., 2019).

# 9. Conclusions and future developments

The main goal of OpenCitations is to provide open scholarly bibliographic and citation data and related services to all possible users, so as to allow anyone to use such data for any purpose. So far, OpenCitations has released more than 450 million citations and has created several interfaces for facilitating their consumption. However, the plan for the next couple of years is to expand the existing citation data available, as well as to create new datasets to serve additional needs.

In particular, following the success of COCI, OpenCitations will release the following new Indexes of existing open citation datasets: WOCI, the OpenCitations Index of open Wikidata citations, DOCI, the OpenCitations Index of open DataCite citations, and DROCI, the OpenCitations Index of open citations within the Dryad Data Repository. This will allow OpenCitations to extend hugely the breadth of coverage of citation data available in its Indexes.

In a major new initiative to be undertaken in collaboration with experts in bibliometrics and scientometrics, OpenCitations will next develop **OpenCitations Meta**, a new database containing additional metadata relating to scholarly publications, which will include many of the kinds of information described in the OpenCitations Data Model but currently lacking in the OpenCitations Corpus, specifically including publications' abstracts, keywords, author affiliations and funding details, information of crucial value for informed bibliometric analyses of research. This information will be of particular interest to universities and research institutes wishing to evaluate the output of their own scholars. In addition, this new database has a crucial role for improving the existing OpenCitations APIs, since it would allow more complex calls on them. It will also reduce the waiting time presently experienced by users, while our systems pause for responses from external API services – e.g. the Crossref API ([https://api.crossref.org](https://api.crossref.org)) – to retrieve the metadata of the bibliographic resources involved in a citation.

By storing such extended bibliographic metadata 'in house' in OpenCitations Meta, we will be able to offer a faster and richer service. Additionally, it will avoid duplication of data by efficiently permitting us to keep in the Meta database a single copy of the metadata for each of the bibliographic entities involved as citing or cited entities in the different OpenCitations' citation indexes, since these same citing and cited entities may be referenced independently within the different Indexes, from Crossref references, Wikidata references, DataCite references, etc.

These developments should be understood as a radical refocusing of OpenCitation's data provision strategy. Initially, the OpenCitations Corpus was conceived as a single database that would contain all our bibliographic and citation information. Now, with the developments (i) of the OpenCitations Indexes containing citation data but not metadata about the citing and cited bibliographic entities, (ii) of the proposed OpenCitations Meta database that will contain bibliographic metadata but not citation data, and (iii) of the new OpenCitations database to be developed as part of our funded Wellcome Trust project that will house textual fragments that constitute the *citation contexts* of each in-text citation occurrence – see Peroni (2018) for additional information –, we are moving to a federated system of interoperable and complementary OpenCitations triplestore resources – following the guidelines in COAR WG Next Generation Repositories (2017).

This is because, as OpenCitations' coverage of the global citation landscape expands, it will become technical inappropriate to handle and maintain everything (metadata, citations, reference lists, in-text reference contexts, abstracts, etc.) within a single repository. Better to organize each specific data type within one of a set of complementary and interoperable repositories, each encoding data in RDF according to the (expanded) OpenCitations Data Model, and all searchable using federated SPARQL queries.

This segregation of distinct data types into different triplestore repositories can then, in future, be extended. And these different repositories can, if necessary, be maintained on different computers at different locations across the Internet or in the cloud, equipped with different hardware according to the specific needs of each repository. General interoperability will be guaranteed by means of SPARQL and its federated service for queries, by using the same generic data model (i.e. the OpenCitations Data Model), and by employing standard Web and Semantic Web protocols and standards for describing all these data.

In future, the OpenCitations "New Corpus" will thus be a set of federated SPARQL-based and OCDM-encoded repositories, each describing a specific type of data, that can talk with each other.

Our plan is to continue to use the original OpenCitations Corpus database itself as a kind of experimental sandbox (https://en.wikipedia.org/wiki/Sandbox_(software_development)) in which all the data types handled by OpenCitations can be stored together, and over which we can test new software and new extensions to the OpenCitations Data Model on a known large but finite set of meaningful papers and their references harvested primarily from the OA subset of Europe PubMed Central.

This approach of using a set of complementary and interoperable repositories searchable using federated SPARQL queries opens the possibility of wider information federation between those resources maintained by OpenCitations and similar interoperable open resources maintained by third parties. Those might provide, for example, information about the publication types of journal items (research articles, comment and opinion pieces, corrigenda, reviews, etc.) in one repository that might be maintained by Crossref; about authors, their ORCiD and/or the Virtual International Authority File (VIAF, https://viaf.org) identifiers and current and past institutional affiliations in another repository, maintained hopefully by ORCiD and VIAF themselves; and about (for example) the geographical focus of published infectious disease reports in yet a third, possibly maintained by WHO. Those resources would be maintained by third parties, thereby spreading the load of providing open scholarly information, while what would unite them would be their use of semantic web technologies, SPARQL, and the common data model.

## Competing Interest Statement

The authors are the Directors of OpenCitations, the subject of this paper.

## References

Bilder, G., Lin, J., & Neylon, C. (2015). Principles for Open Scholarly Infrastructures. Figshare. https://doi.org/10.6084/m9.figshare.1314859

Bizer, C., Heath, T., & Berners-Lee, T. (2009). Linked Data—The Story So Far. International Journal on Semantic Web and Information Systems, 5(3), 1–22. https://doi.org/10.4018/jswis.2009081901

COAR WG Next Generation Repositories. (2017). Behaviours and Technical Recommendations of the COAR Next Generation Repositories Working Group [Recommendation]. Retrieved from COAR website: https://www.coar-repositories.org/files/NGR-Final-Formatted-Report-cc.pdf

Corbatto, M., & Dattolo, A. (2018). A Web Application for Creating and Sharing Visual Bibliographies. In A. González-Beltrán, F. Osborne, S. Peroni, & S. Vahdati (Eds.), Semantics, Analytics, Visualization (pp. 78–94). https://doi.org/10.1007/978-3-030-01379-0_6

Cyganiak, R., Wood, D., & Krötzsch, M. (2014). RDF 1.1 Concepts and Abstract Syntax [W3C Recommendation]. Retrieved from World Wide Web Consortium website: https://www.w3.org/TR/rdf11-concepts/

Daquino, M., Peroni, S., & Shotton, D. (2019). The OpenCitations Data Model. Figshare. https://doi.org/10.6084/m9.figshare.3443876

Di Iorio, A., Peroni, S., & Poggi, F. (2019). Open data to evaluate academic researchers: An experiment with the Italian Scientific Habilitation. Proceedings of the 17th International Conference on Scientometrics and Informetrics (ISSI 2019). http://arxiv.org/abs/1902.03287

Ferguson, C., McEntrye, J., Bunakov, V., Lambert, S., van der Sandt, S., Kotarski, R., … McCafferty, S. (2018). Survey of Current PID Services Landscape (Deliverable No. D3.1). Retrieved from FREYA project (EC Grant Agreement No 777523) website: https://www.project-freya.eu/en/deliverables/freya_d3-1.pdf

Harris, S., & Seaborne, A. (2013). SPARQL 1.1 Query Language [W3C Recommendation]. Retrieved from World Wide Web Consortium website: https://www.w3.org/TR/sparql11-query/

Heibi, I., Peroni, S., & Shotton, D. (2019a). COCI, the OpenCitations Index of Crossref open DOI-to-DOI citations. Scientometrics, 121 (2), 1213–1228. https://doi.org/10.1007/s11192-019-03217-6

Heibi, I., Peroni, S., & Shotton, D. (2019b). Crowdsourcing open citations with CROCI - An analysis of the current status of open citations, and a proposal. Proceedings of the 17th International Conference on Scientometrics and Informetrics (ISSI 2019). http://arxiv.org/abs/1902.02534

Heibi, I., Peroni, S., & Shotton, D. (2019c). Enabling text search on SPARQL endpoints through OSCAR. Data Science, 1–23. https://doi.org/10.3233/DS-190016

Hosseini, A., Ghavimi, B., Boukhers, Z., & Mayr, P. (2019). EXCITE - A toolchain to extract, match and publish open literature references. Presented at the 19th ACM/IEEE on Joint Conference on Digital Libraries, Urbana-Champaign, Illinois, USA. Retrieved from https://philippmayr.github.io/papers/JCDL2019-EXCITE-demo.pdf

Hutchins, B. I., Baker, K. L., Davis, M. T., Diwersy, M. A., Haque, E., Harriman, R. M., … Santangelo, G. M. (2019). The NIH Open Citation Collection: A public access, broad coverage resource. PLOS Biology, 17(10), e3000385. https://doi.org/10.1371/journal.pbio.3000385

Kamińska, A. M. (2017). Plos One – A Case Study Of Citation Analysis Of Research Papers Based On The Data In An Open Citation Index (The Opencitations Corpus). Zenodo. https://doi.org/10.5281/zenodo.1066316

Lauscher, A., Eckert, K., Galke, L., Scherp, A., Rizvi, S. T. R., Ahmed, S., … Klein, A. (2018). Linked Open Citation Database: Enabling Libraries to Contribute to an Open and Interconnected Citation Graph. Proceedings of the 18th ACM/IEEE on Joint Conference on Digital Libraries - JCDL '18, 109–118. https://doi.org/10.1145/3197026.3197050

Lebo, T., Sahoo, S., & McGuinness, D. (2013). PROV-O: The PROV Ontology. W3C Recommendation. World Wide Web Consortium. https://www.w3.org/TR/prov-o/ (last visited 14 September 2019)

Moreau, L., & Missier, P. (2013). PROV-DM: The PROV Data Model. W3C Recommendation. World Wide Web Consortium. http://www.w3.org/TR/prov-dm/ (last visited 14 September 2019)

OpenCitations. (2019). Data about website accesses and Figshare views and downloads from April 2018 to March 2019. Figshare. https://doi.org/10.6084/m9.figshare.8050352.v2

Pearson, S. (2019). Using open citation data to identify new research opportunities. Retrieved June 22, 2019, from Library Research Plus website: https://blog.research-plus.library.manchester.ac.uk/2019/03/04/using-open-citation-data-to-identify-new-research-opportunities/

Peroni, S. (2018). The Wellcome Trust funds OpenCitations. Retrieved June 26, 2019, from OpenCitations blog website: https://opencitations.wordpress.com/2018/12/23/the-wellcome-trust-funds-opencitations/

Peroni, S., & Shotton, D. (2018a). Open Citation: Definition. Figshare. https://doi.org/10.6084/m9.figshare.6683855

Peroni, S., & Shotton, D. (2018c). The SPAR Ontologies. In L. Rutkowski, R. Scherer, M. Korytkowski, W. Pedrycz, R. Tadeusiewicz, & J. M. Zurada (Eds.), The Semantic Web – ISWC 2018: 17th International Semantic Web Conference, Monterey, CA, USA, October 8–12, 2018, Proceedings, Part II (pp. 119–136). https://doi.org/10.1007/978-3-030-00668-6_8

Peroni, S., & Shotton, D. (2019). Open Citation Identifier: Definition. Figshare. https://doi.org/10.6084/m9.figshare.7127816

Peroni, S., Shotton, D., & Vitali, F. (2016). A document-inspired way for tracking changes of RDF data—The case of the OpenCitations Corpus. In L. Hollink, S. Darányi, A. Meroño Peñuela, & E. Kontopoulos (Eds.), Detection, Representation and Management of Concept Drift in Linked Open Data (Vol. 1799, pp. 26–33). http://ceur-ws.org/Vol-1799/Drift-a-LOD2016_paper_4.pdf

Peroni, S., Shotton, D., & Vitali, F. (2017). One Year of the OpenCitations Corpus. In C. d'Amato, M. Fernandez, V. Tamma, F. Lecue, P. Cudré-Mauroux, J. Sequeda, … J. Heflin (Eds.), The Semantic Web – ISWC 2017 (pp. 184–192). https://doi.org/10.1007/978-3-319-68204-4_19

Shotton, D. (2013a). Linked Data 101. Retrieved September 14, 2019, from OpenCitations blog website: https://semanticpublishing.wordpress.com/2013/03/01/linked-data-101/

Shotton, D. (2013b). Open citations. Nature 502: 295–297. DOI: https://doi.org/10.1038/502295a

Shotton, D. (2018). Citations as First-Class Data Entities: Introduction. Retrieved June 22, 2019, from OpenCitations blog website: https://opencitations.wordpress.com/2018/02/19/citations-as-first-class-data-entities-introduction/

Sugimoto, C. R., Waltman, L., Larivière, V., van Eck, N. J, Boyack, K. W., Wouters, P., & de Rijcke, S. (2017). Open citations: A letter from the scientometric community to scholarly publishers. ISSI Society. http://issi-society.org/open-citations-letter (last visited 23 March 2019)

Teufel, S., Siddharthan, A., & Tidhar, D. (2006). Automatic classification of citation function. Proceedings of the 2006 Conference on Empirical Methods in Natural Language Processing (EMNLP '06), 103. https://doi.org/10.3115/1610075.1610091

van Eck, N. J., & Waltman, L. (2010). Software survey: VOSviewer, a computer program for bibliometric mapping. Scientometrics, 84 (2), 523–538. https://doi.org/10.1007/s11192-009-0146-3

Wilkinson, M. D., Dumontier, M., Aalbersberg, Ij. J., Appleton, G., Axton, M., Baak, A., … Mons, B. (2016). The FAIR Guiding Principles for scientific data management and stewardship. Scientific Data, 3, 160018. https://doi.org/10.1038/sdata.2016.18

Zhu, Y., Yan, E., Peroni, S., & Che, C. (2019). Nine Million Books and Eleven Million Citations: A Study of Book-Based Scholarly Communication Using OpenCitations. arXiv. http://arxiv.org/abs/1906.06039